\documentclass[aps,prb,twocolumn,groupedaddress,showpacs]{revtex4}
\usepackage{graphicx}
\usepackage{amsmath,amssymb,amsfonts}
\usepackage{natbib}

\newcommand{\bd}{\begin{displaymath}}
\newcommand{\ed}{\end{displaymath}}
\newcommand{\be}{\begin{equation}}
\newcommand{\ee}{\end{equation}}
\newcommand{\bs}{\begin{subequations}}
\newcommand{\es}{\end{subequations}}
\newcommand{\ba}{\begin{eqnarray}}
\newcommand{\ea}{\end{eqnarray}}

\begin{document}

\title{A generalized Chudley-Elliott vibration-jump model\\
in activated atom surface diffusion}

\author{R. Mart\'{\i}nez-Casado$^{a,b}$}
\email{ruth@imaff.cfmac.csic.es}

\author{J. L. Vega$^b$}
\email{jlvega@imaff.cfmac.csic.es}

\author{A. S. Sanz$^b$}
\email{asanz@imaff.cfmac.csic.es}

\author{S. Miret-Art\'es$^b$}
\email{s.miret@imaff.cfmac.csic.es}

\affiliation{$^a$Lehrstuhl f\"ur Physikalische Chemie I,
Ruhr-Universit\"at Bochum, D-44801 Bochum, Germany}

\affiliation{$^b$Instituto de Matem\'aticas y F\'{\i}sica Fundamental,
Consejo Superior de Investigaciones Cient\'{\i}ficas,
Serrano 123, 28006 Madrid, Spain}

\date{\today}

\begin{abstract}
Here the authors provide a generalized Chudley-Elliott expression for
activated atom surface diffusion which takes into account the coupling
between both low-frequency vibrational motion (namely, the frustrated
translational modes) and diffusion.
This expression is derived within the Gaussian approximation framework
for the intermediate scattering function at low coverage.
Moreover, inelastic contributions (arising from creation and
annihilation processes) to the full width at half maximum of the
quasi-elastic peak are also obtained.
\end{abstract}


\maketitle


\section{Introduction}
\label{sec1}

In 1960 Chudley and Elliott\cite{elliott} proposed a jump diffusion
model for neutron scattering from a liquid, where the latter was
assumed to be locally a quasi-crystalline structure.
Within this model, the diffusive motion undergone by the (liquid)
atoms is described in terms of large discrete jumps; otherwise, the
atoms remain oscillating around their fixed (lattice) point between
jumps.
Because of the very different timescales ruling each type of motion
(diffusive and oscillatory), they are considered as completely
uncorrelated.
This jump model has also been extensively used to interpret the full
width at half maximum (FWHM) of the so-called {\it quasi-elastic
peak} ($Q$ peak) observed in quasi-elastic helium atom (surface)
scattering (QHAS) experiments.\cite{toennies,graham,vega,raul,eli}
This peak, obtained from time-of-flight measurements converted to
energy transfers, accounts for the adatom surface diffusion process.
Apart from the $Q$ peak, in this type of experiments one also observes
oscillatory motions associated to the temporary trapping of the adatom
inside surface potential wells.
This is the so-called {\it frustrated translational mode} (or
$T$ mode), the lowest frequency mode of the adsorbate, whose
lineshape is located very close to the $Q$ peak which is centered
at zero energy transfer.

Following the Chudley-Elliot model, the coupling between the diffusive
and vibrational motions is very often neglected in theoretical
descriptions.
Of course, for short parallel momentum transfers of the scattered
particles (He atoms), the coupling between both motions can be
naturally disregarded, since no overlapping between the corresponding
peaks will be observed.
However, the same does not hold for large momentum transfers.
This is important because the experimental deconvolution of the $Q$ and
$T$-mode peaks is often carried out independently.\cite{toennies,graham}
In this paper, we propose a working formula for the whole lineshape
based on the Gaussian approximation and that takes into account a
generalized Chudley-Elliott expression dealing with the coupling
between the diffusive peak and the low-frequency vibrational modes.
In this way, the experimental deconvolution procedure can be carried
out in a more appropriate manner, and the typical information extracted
(jump distributions, and diffusion and friction coefficients) from both
motions at the same time would lead to more realistic values.


\section{Theory}
\label{sec2}

In real diffusion experiments performed by means of the QHAS technique,
the observable magnitude is the {\it differential reflection
probability},\cite{vanHove}
\be
 \frac{d^2 {\mathcal R} (\Delta {\bf K}, \omega)}{d\Omega d\omega}
  = n_d F S(\Delta {\bf K}, \omega) .
 \label{eq1}
\ee
This magnitude gives the probability that the probe (He) atoms
scattered from the diffusing collective (chattered on the surface)
reach a certain solid angle $\Omega$ with an energy exchange
$\hbar \omega = E_f - E_i$ and parallel momentum transfer
$\Delta {\bf K} = {\bf K}_f - {\bf K}_i$.
In Eq.~(\ref{eq1}), $n_d$ is the (diffusing) particle concentration,
and $F$ is the {\it atomic form factor}, which depends on the
interaction potential between the probe atoms in the beam and the
adparticles.
The function $S(\Delta {\bf K},\omega)$ is the so-called
{\it dynamic structure factor}, which provides a complete information
about the dynamics and structure of the adsorbed particle ensemble,
and therefore also about the surface diffusion process and vibrational
spectroscopy of their low frequency modes.
Information about long distance correlations is obtained when using
small values of $\Delta {\bf K}$, while long-time scale correlation
information is available at small energy transfers, $\hbar \omega$.

In a series of previous works,\cite{vega,raul,eli} we have analyzed in
detail the lineshape structure of both the $Q$ and $T$-mode peaks in
terms of the standard jump Chudley-Elliot model.
This model can be improved in order to provide a better analysis of
the experimental lineshapes.
For this goal, we start expressing the dynamic structure factor that
appears in Eq.~(\ref{eq1}) as
\be
 S(\Delta {\bf K},\omega) =
  \frac{1}{2 \pi} \int e^{-i\omega t} \ \! I(\Delta{\bf K},t) \ \! dt ,
 \label{eq2}
\ee
where
\be
 I(\Delta {\bf K},t) \equiv
  \langle e^{-i\Delta {\bf K} \cdot
   [{\bf R}(t) - {\bf R}(0)] } \rangle
  = \langle e^{-i \Delta K \!
    \int_0^t v_{\Delta {\bf K}} (t') \ \! dt'} \rangle
 \label{eq3}
\ee
is the so-called {\it intermediate scattering function} (the brackets
denote ensemble averaging), with ${\bf R} (t)$ the trajectory of the
adparticles and $v_{\Delta {\bf K}}$ the corresponding velocity
projected onto the direction of the parallel momentum transfer
$\Delta {\bf K}$ (with length $\Delta K = \|\Delta {\bf K}\|$).
Both the dynamic structure factor and the intermediate scattering
function can be readily obtained from standard Langevin numerical
simulations.\cite{gardiner}
Note that this type of simulation is valid as long as no collective,
clustering, or self-organizing behaviors are present; in such cases
one should make use of more detailed numerical techniques (e.g.,
molecular dynamics), which allow us to follow the processes that happen
during short timescales (as the aforementioned ones).

After a second order cumulant expansion in $\Delta {\bf K}$ in the
r.h.s.\ of the second equality of Eq.~(\ref{eq3}), $I(\Delta K, t)$
can be expressed as
\be
 I(\Delta K, t) \approx
  e^{- \Delta K^2 \int_0^t (t - t') \mathcal{C}(t') dt'} ,
 \label{eq4}
\ee
where $\mathcal{C}(t) = \langle v_{\Delta {\bf K}} (0) v_{\Delta
{\bf K}}(t) \rangle $ is the {\it velocity autocorrelation function}.
This is the so-called {\it Gaussian approximation},\cite{mcquarrie,kubo,yip}
which is exact provided velocity correlations at more than two
different times are negligible.\cite{mcquarrie}
If non-Gaussian corrections are considered the theory cannot be
expressed in a simple analytical manner; deviations from the Gaussian
approximation are a clear indicator of the importance of such
corrections.

At low coverage, adsorbate-adsorbate interactions can be neglected;
diffusion is then fully characterized by only studying the dynamics of
an isolated adsorbate on the surface.
This is the so-called {\it single adsorbate approximation} in
self-diffusion.
Within the Langevin framework, the adsorbate-substrate interaction
is described in terms of two contributions: (i) a deterministic,
fitted adiabatic potential, $V$, which models the adsorbate-surface
interaction at zero surface temperature $T_s = 0$~K, and leading
therefore to a deterministic force, $F = -\nabla V$; and (ii) a
stochastic force (Gaussian white noise) accounting for the vibrational
effects induced by the temperature on the (surface) lattice atoms that
act on the adsorbate.

For an almost flat surface ($V \approx 0$), any direction is equivalent
and therefore the dimensionality of the numerical Langevin simulation
reduces to one.
The corresponding numerical velocity autocorrelation function (in
self-diffusion) then follows an exponential decreasing behavior,
\be
 \mathcal{C}(t) = \frac{k_B T_s}{m} \ \! e^{- \gamma t} ,
 \label{eq5}
\ee
where $m$ is the adparticle mass, $k_B$ the Boltzmann constant, and
$\gamma$ the friction coefficient, which is related to the Gaussian
white noise source through the fluctuation-dissipation theorem.
Introducing (\ref{eq5}) into Eq.~(\ref{eq4}), one obtains\cite{kubo}
\be
 I(\Delta K, t) = \exp \left[- \chi^2
   \left( e^{- \gamma t} + \gamma t - 1 \right) \right] ,
 \label{eq6}
\ee
where $\chi$ is the  {\it shape parameter}, defined as\cite{ruth}
\be
 \chi \equiv \sqrt{\langle v^2 \rangle} \ \Delta K  / \gamma
  = \bar{l} \ \Delta K .
 \label{eq7}
\ee
Here, $\sqrt{\langle v^2 \rangle} = \sqrt{k_B T_s/m}$ is the thermal
velocity in one dimension, and $\bar{l} \equiv  \sqrt{ \langle v^2
\rangle} / \gamma$ is the mean free path.
The diffusion coefficient is related to the friction as
$D \equiv \langle v^2 \rangle / \gamma$.
The only information about the structure of the lattice is found in
the shape parameter [Eq.~(\ref{eq7})] through $\Delta K$.
When large parallel momentum transfers are under consideration, both
the periodicity and structure of the surface have to be taken into
account.
Consequently, the shape parameter should be changed for different
lattices.

The simplest model including the periodicity of the surface is
that developed by Chudley and Elliott,\cite{elliott} who proposed
a master equation for the pair-distribution function in space and
time\cite{vanHove} assuming instantaneous discrete jumps on a
two-dimensional Bravais lattice. The Fourier transform of this master
equation gives rise to an exponential function for the intermediate
scattering function,
\be
 I(\Delta {\bf K}, t) =
  I(\Delta {\bf K}, 0) \ \! e^{- \Gamma_{\nu} (\Delta {\bf K}) \ \! t} .
 \label{eq8}
\ee
The inverse of the correlation time, $\Gamma_{\nu} (\Delta {\bf K})$,
has a periodic dependence on $\Delta {\bf K}$ of the form
\be
 \Gamma_{\nu} (\Delta {\bf K}) = \nu
 \sum_{\bf j} P_{\bf j} \ \! [1 - \cos({\bf j} \cdot \Delta {\bf K})] ,
 \label{eq9}
\ee
where $\nu$ is the total jump rate out of an adsorption site, and
$P_{\bf j}$ is the relative probability that a jump with a displacement
vector ${\bf j}$ occurs.
The time Fourier transform of Eq.~(\ref{eq8}) is a Lorenztian
function with FWHM equal to $\Gamma_{\nu} (\Delta {\bf K})$.
At sufficiently small values of the parallel momentum transfer,
$\Gamma_{\nu} (\Delta {\bf K})$ displays a cuadratic dependence on
$\chi$ according to Eq.~(\ref{eq7}),
\be
 \Gamma_{\nu} (\Delta {\bf K}) \approx 2 D \Delta {\bf K}^2
  = 2 \gamma \chi^2 .
 \label{eq10}
\ee
From Eqs.~(\ref{eq9}) and (\ref{eq10}) the shape parameter can be now
generalized in terms of the jump distribution on the surface Bravais
lattice, defining
\be
 \chi_l (\Delta {\bf K}) \equiv \sqrt{\frac{\Gamma_{\nu}
 (\Delta {\bf K})}{2 \gamma}} ,
 \label{eq11}
\ee
which is valid for any value of the parallel momentum transfer, and
where the subscript $l$ denotes the importance of the lattice.
Note that, as a consequence of the effects induced by the periodicity
and structure of the surface Bravais lattice, the shape parameter
$\chi$ for free diffusion is replaced by $\chi_l$.
This is related to the exponential dependence on time displayed by
the intermediate scattering function.
Equation~(\ref{eq11}) can then be seen as a definition for the new
shape parameter, $\chi_l$, which takes into account the surface
structure.

Numerical Langevin simulations for corrugated surface potentials have
shown\cite{vega} that the Gaussian approximation is quite often
appropriate.
In such cases, the velocity autocorrelation function can be fitted to
a general function described by
\be
 \mathcal{C}(t) = \langle v^2_{\Delta {\bf K}} \rangle
  \ \! e^{- \gamma t} \ \! \cos (\omega_T \ \! t + \delta) ,
 \label{eq12}
\ee
where $\omega_T$ gives the $T$-mode frequency (including the shift due
to the surface temperature) and $\delta$ is a phase which is added to
better fit numerical results.
The $\omega_T$ frequency is related to the harmonic one, $\omega_0$,
according to the relation
\be
 \omega_T = \sqrt{\omega_0^2 - \frac{\gamma^2}{4}} .
 \label{eqT}
\ee
Analogously, in the case of a harmonic oscillator, $\delta$ can also
be expressed\cite{ruth2} in terms of a relationship between $\gamma$
and $\omega_T$ [in particular,
$\delta = (\tan)^{-1} (\gamma/2\omega_T$)].
However, this constraint is lost when we apply Eq.~(\ref{eq12}) to
cases of interest due to anharmonicities; as seen below, in such cases,
$\gamma$, $\omega_T$, and $\delta$ are considered as fitting parameters.

Introducing (\ref{eq12}) into Eq.~(\ref{eq4}) the intermediate
scattering function can be expressed in terms of a double infinite
sum of exponential functions, each contribution coming from the
different quantum states of the $T$-mode oscillator.
This allows to use $\chi_l$ instead of $\chi$, and, in analogy to
the result obtained in Ref.~\onlinecite{vega}, the intermediate
scattering function of the full process can be generalized as
\begin{multline}
 I({\bf K},t) = \exp{[- \chi_l^2 \, f(\omega_T,\delta,t)]} \\
 = e^{- \chi_l^2 A_1 - \chi_l^2 A_2 t} \sum_{n,m}^{\infty}
  \frac{(-1)^n (-1)^m}{n! m!} \chi_l^{2(n+m)} A_3^n A_4^m \\
  \times
   e^{-i (m-n) \delta} e^{- (m+n) \gamma t} e^{-i (m-n) \omega_T t} ,
 \label{eq13}
\end{multline}
%
where
\bs
\ba
 A_1 & = & \frac{\gamma^2 [2 \gamma \omega_T \sin \delta +
  (\omega_T^2 - \gamma^2)\cos \delta]}{(\gamma^2 + \omega_T^2)^2} ,
 \\
 A_2 & = & \frac{\gamma^2 ( \gamma \cos \delta - \omega_T \sin \delta)}
  {\gamma^2 + \omega_T^2} ,
 \label{eq14b}
\\
 A_3 & = & \frac{\gamma^2}{2 (\gamma - i \omega_T)^2} ,
 \\
 A_4 & = & \frac{\gamma^2}{2  (\gamma + i \omega_T)^2} .
 \ea
\es
Now, Fourier transforming Eq.~(\ref{eq13}), an analytical expression
for the dynamic structure factor is also readily obtained,
\begin{multline}
 S({\Delta \bf K},\omega) = \\
 \frac{e^{-\chi_l^2 A_1}}{\pi} \sum_{n,m=0}^{\infty}
  \frac{(-1)^{n+m} \chi_l^{2(n+m)} A_3^n A_4^m}{n! m!}
  \ \! e^{-i (m-n) \delta} \\
 \times\frac{[ \chi_l^2 A_2 + (n+m) \gamma]}
  {\ [\omega -(n-m) \omega_T]^2 + [\chi_l^2 A_2 + (n+m) \gamma]^2} ,
\label{eq15}
\end{multline}
%
where the $n$ and $m$ indexes represent different excitations of
$T$-mode creation and annihilation events.

The double sum over Lorentzian shapes in Eq.~(\ref{eq15}) clearly
shows that both motions (diffusion and oscillations) cannot be
separated at all.
Only the terms with $m=n$ (events with a zero energy balance)
contribute to the $Q$ peak, while those with $n \neq m$ will contribute
to the $T$-mode peaks.
As discussed in Refs.~\onlinecite{vega} and \onlinecite{ruth}, the
shape parameter also governs the shape of the peaks.
Depending on the new shape parameter (the dependence on $\Delta {\bf K}$ is
only through $\chi_l$), the sums will contribute globally to different
intermediate shapes ranging from a Gaussian function (high values of $\chi_l$)
to a Lorentzian one ($\chi_l \ll 1$).
This smooth shape transition is the so-called {\it motional narrowing effect},
well-known in the theory of nuclear magnetic resonance
lineshapes.\cite{vleck} For a given system, this effect could be
observable by changing the $\Delta {\bf K}$ value, as shown below.
In Ref.~\onlinecite{vega} we proposed a similar functional form to
Eq.~(\ref{eq15}) as a working formula, but using instead the shape
parameter for a flat surface [$\chi$, as given by Eq.~(\ref{eq7})].
With the new expression provided here, where $\chi$ has been replaced
by $\chi_l$, one can now determine the lineshapes displayed by the $Q$
and $T$-mode peaks in numerical calculations or in experiments after
deconvolution for the whole first Brillouin zone.
This is an important issue because the experimental deconvolution
procedure is generally carried out for each peak separately (i.e.,
there is no overlapping), pre-assuming a given shape (an effective
Lorenztian function) with no theoretical justification.
This assumption leads to a proportionality between the effective width
and the diffusion coefficient, which is not correct in general.
In this sense, to perform the whole deconvolution procedure a
generalization of such an equation is convenient, which will allow
to extract, at the same time, good diffusion coefficients, information
about jump mechanisms, friction coefficients, and low $T$-mode
frequencies.
When more than one Lorentzian function contribute to Eq.~(\ref{eq15}),
diffusion coefficients should be then obtained from the calculated
jump distributions, and therefore not directly extracted from the
experimental $Q$ peak.

From the generalized expression for the lineshape of the full process,
given by Eq.~(\ref{eq15}), the width of each sum contributing to the
$Q$ peak reads now as
\be
 \Gamma_n^Q (\Delta {\bf K}) =
  \frac{\ \! \Gamma_{\nu} (\Delta {\bf K})}{2 \gamma}
   \ \! A_2 + 2 n \gamma .
 \label{eq16}
\ee
Note that $A_2$ provides information about the dynamical process
through the frequency associated to the $T$-mode, the friction
coefficient and the phase of the
velocity autocorrelation function [see Eqs.~(\ref{eq12}) and
(\ref{eq14b})].
The surface temperature dependence appears implicitly in the shift
of the position of the $T$-mode frequency and the jump distribution
$P_{\bf j}$ through $\Gamma_{\nu} (\Delta {\bf K})$ according to
Eq.~(\ref{eq9}).
Typically, the most important contribution to the $Q$ peak at low
temperatures is $\Gamma_{n=0}^Q$.
Within this model, if  $\delta = 0$, $\omega_T = 0$, and the regime
of very small values of the parallel momentum transfer is studied,
Eq.~(\ref{eq10}) is then satisfied and $\Gamma_0^Q (\Delta {\bf K})
= D \Delta {\bf K}^2$, as obtained when a pure (or effective)
Lorentzian function is assumed.
Inelastic contributions (with zero energy balance),
$\Gamma_{n \neq 0}^Q$, to the total width due to creation and
annihilation processes of the $T$-mode are important with the surface
temperature and increases by integer numbers of $\gamma$.
Mixed Lorentzian-Gaussian lineshapes are very often obtained, the
FWHM being calculated from the general expression of the dynamic
structure factor.

In a similar fashion, the width of each sum contributing to the
$T$-peaks, according to Eq.~(\ref{eq15}), is given by
\be
 \Gamma_{n+m}^T (\Delta {\bf K}) =
  \frac{\ \! \Gamma_{\nu} (\Delta {\bf K})}{2 \gamma}
   \ \! A_2 + (n+m) \gamma ,
 \label{eq17}
\ee
where the jump distribution $P_{\bf j}$ is also involved in the
vibrational relaxation process through $\Gamma_{\nu} (\Delta {\bf K})$.
It is generally assumed that the corresponding width is $\gamma$,
which (within this model) is obtained when $n=1$, $m=0$, $\delta = 0$,
and $\gamma \ll \omega_0$; note that the first term in Eq.~(\ref{eq17})
behaves as $\gamma^2 / \omega_0^2$, which is much smaller than the
second term (which is equal to $\gamma$).
In this case, the lineshape is a pure Lorentzian function.


\section{Results}
\label{sec3}

\begin{figure}
 \includegraphics[width=7cm]{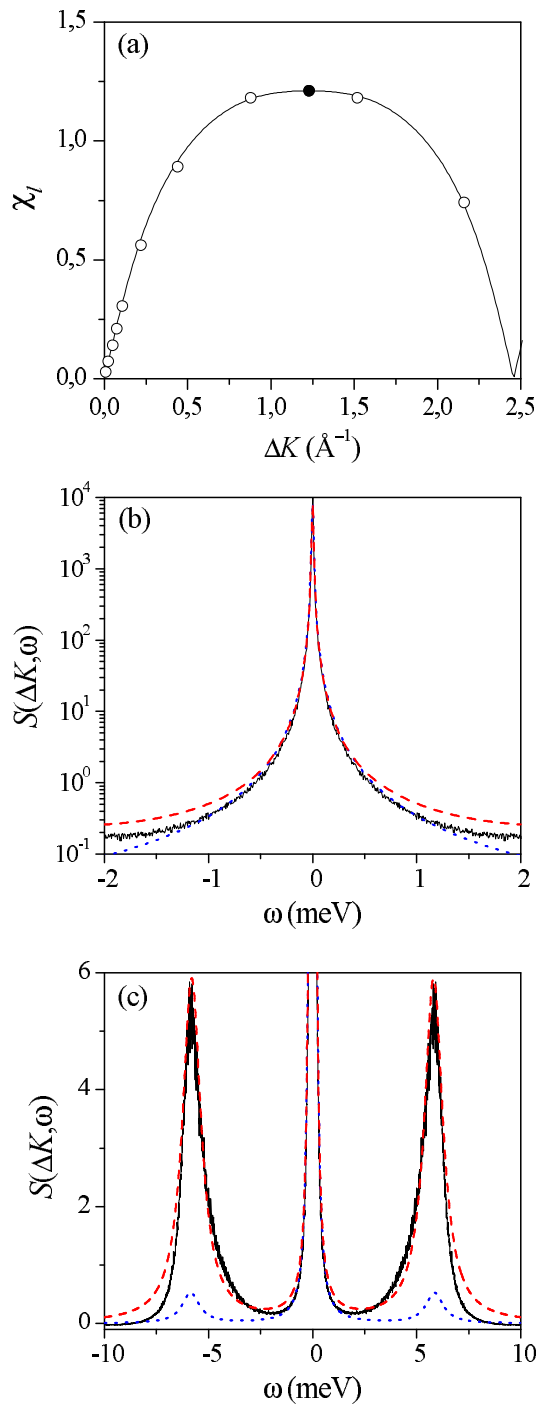}
 \caption{\label{fig1}
  (Color online.)
  (a) Shape parameter $\chi_l$ as a function of the parallel momentum
  transfer covering the first Brillouin zone for the Na/Cu(001)
  system at $T_s = 150$~K.
  Open circles correspond to the fitting of the numerical simulation,
  while the solid line is the result obtained from Eq.~(\ref{eq11}).
  The dynamic structure factor for the maximum momentum
  transfer, $\Delta K = 1.23$~\AA$^{-1}$ (solid circle), is plotted in
  panels for the $Q$ peak (b) and for the two $T$-mode peaks (c).
  Solid (black) lines corresponds to the numerical simulation; dotted
  (blue) lines represent the result obtained from using the working
  formula with no fitting parameters; dashed (red) lines are the result
  when three fitting parameters ($\gamma$, $\omega_T$, and $\delta$)
  are employed.}
\end{figure}

As an illustration of the theoretical approach proposed and described
in Sec.~\ref{sec2}, we have carried out numerical Langevin simulations
for the Na/Cu(001) system, which has been largely analyzed both
experimentally and theoretically.
The adiabatic nonseparable adsorbate-substrate interaction is taken
from Ref.~\onlinecite{graham}, with a coverage of 2.8$\%$ and for a
surface temperature $T_s = 150$~K.
In Fig.~\ref{fig1}(a), the shape parameter $\chi_l$ is plotted
as a function of the parallel momentum transfer covering the first
Brillouin zone.
The solid curve has been obtained from Eqs.~(\ref{eq9}) and
(\ref{eq11}) after jump distributions were calculated.
The circles give the fittings of the different Langevin simulations
to the dynamic structure factor given by Eq.~(\ref{eq15}).
In panels (b) and (c) of Fig.~\ref{fig1}, the quality of the fitting
for $\Delta K = 1.23$~\AA$^{-1}$ [black circle in panel (a)], for the
$Q$ and the two $T$-modes peaks, respectively, are shown; dotted lines
correspond to the application of Eq.~(\ref{eq15}) with no adjustable
parameters (nominal values: $\gamma = 0.6$~meV, $\omega_T= 5.99$~meV,
and $\delta =0.05$), and dashed lines to the case where
Eq.~(\ref{eq15}) has been used as a working formula with three
different adjustable parameters ($\gamma = 0.53$~meV,
$\omega_T = 5.82$~meV, and $\delta =0.032$).
The whole spectrum is better reproduced in the last case; in the first
fitting the right positions of the $T$-mode peaks are obtained but with
very small intensities compared to the numerical simulations.
Note, however, that for the $Q$ peak, both fittings are approximately
the same.
The small deviations from the nominal values of the three parameters
are a clear indicator of the non-Gaussian behavior of the dynamical
system.
Concerning the shape of the $Q$ peak, when $\chi_l$ is greater than one
(central region of the Brillouin zone), a mixed Gaussian-Lorentzian
function is expected.

\begin{figure}
 \includegraphics[width=7cm]{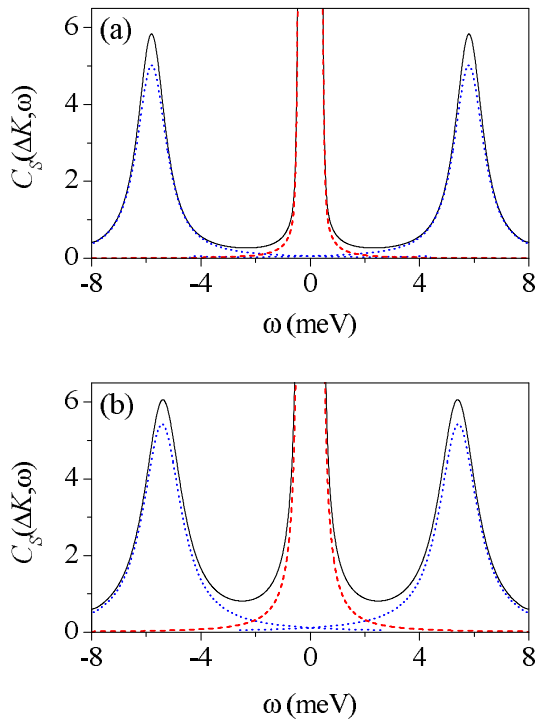}
 \caption{\label{fig2}
  (Color online.)
  Numerical convolution (solid line) of the $Q$ and $T$-mode peaks for
  the Na/Cu(001) system with parallel momentum transfer $\Delta K =
  1.23$~\AA$^{-1}$ at two different surface temperatures: (a) 150~K
  and (b) 250~K.
  Dashed (red) lines correspond to the convolution of a pure Lorentzian
  function for the $Q$ peak; dotted (blue) lines denote the same, but
  for the $T$-mode peaks.}
\end{figure}

In Fig.~\ref{fig2}, the convolution of the dynamic structure factor
(denoted by $C_S$) obtained from the Langevin numerical simulation
for $\Delta K = 1.23$~\AA$^{-1}$ [see Fig.~\ref{fig1}(c)] is plotted
(solid line).
Moreover, the convolution of separate Lorentzian functions for the
$Q$ and $T$-mode peaks are also shown (dashed line for the $Q$ peak and
dotted lines for the $T$-mode peaks).
The two panels correspond to two different surface temperatures:
(a) 150~K and (b) 250~K.
The response function of the experimental setup is known and given in
Ref.~\onlinecite{graham}.
Discrepancies between the standard procedure and the one proposed here
are apparent, as indicated by their different backgrounds, widths, and
peak intensities.

As a word of conclusion, we consider that a convolution of the whole
spectrum should be carried out in order to properly take into account
the contributions of the events with zero energy balance in the
$Q$ peak, as well as the existing overlapping between both $Q$ and
$T$-mode peaks at larger momentum transfers and high surface
temperatures.


\section*{Acknowledgements}

This work was supported in part by DGCYT (Spain), Project No.\
FIS2004-02461.
One of the authors \mbox{(R.M.-C.)} would like to thank the University of
Bochum for support from the Deutsche Forschungsgemeinschaft, contract
No.\ SFB 558.
Two of the authors (J.L.V.\ and A.S.S.) would like to thank the
Ministerio de Educaci\'on y Ciencia (Spain) for a predoctoral grant
and a ``Juan de la Cierva'' Contract, respectively.


\end{document}